\documentclass[10pt]{article}

\def\CRAS{C.~R.~Acad.~Sc.~Paris}

\def\cKdV{c-KdV}

\def\Kaone {K_{1,{\rm a}}}

\def\KFIVone {K_{1,{\rm IV}}}
\def\KFIVtwo {K_{2,{\rm IV}}}

\def\KFVIone {K_{1,{\rm VI}}}

\newcommand{\pard}[2]{\frac{\partial #1}{\partial #2}}

\begin{document}

\title{Completeness of the cubic and quartic H\'enon-Heiles 
Hamiltonians\footnote
{Corresponding author RC. Preprint S2004/047. nlin.SI/0507011}
}

\author{Robert CONTE~$^\dag$,
Micheline MUSETTE~$^\ddag$
and
Caroline VERHOEVEN~$^\ddag$
{}
\\ \dag Service de physique de l'\'etat condens\'e (URA 2464), CEA--Saclay
\\ F--91191 Gif-sur-Yvette Cedex, France
\\ E-mail:  Conte@drecam.saclay.cea.fr
{}\\
\\ \ddag Dienst Theoretische Natuurkunde, Vrije Universiteit Brussel
and
\\~~International Solvay Institutes for Physics and Chemistry
\\~~Pleinlaan 2, B--1050 Brussels, Belgium
\\~~E-mail: MMusette@vub.ac.be, CVerhoev@vub.ac.be
}

\maketitle

\hfill 

{\vglue -10.0 truemm}
{\vskip -10.0 truemm}

\begin{abstract}
\noindent
The quartic H\'enon-Heiles Hamiltonian
$H = (P_1^2+P_2^2)/2+(\Omega_1 Q_1^2+\Omega_2 Q_2^2)/2
 +C Q_1^4+ B Q_1^2 Q_2^2 + A Q_2^4
 +(1/2)(\alpha/Q_1^2+\beta/Q_2^2) - \gamma Q_1$
passes the Painlev\'e test for only four sets of values of the constants.
Only one of these, identical to the traveling wave reduction of the Manakov
system, has been explicitly integrated (Wojciechowski, 1985),
while the three others are not yet integrated 
in the generic case $(\alpha,\beta,\gamma)\not=(0,0,0)$.
We integrate them 
by building a birational transformation to two fourth order first degree
equations in the classification (Cosgrove, 2000) of such polynomial equations
which possess the Painlev\'e property.
This transformation involves the stationary reduction
of various partial differential equations (PDEs).
The result is the same as for the three cubic H\'enon-Heiles Hamiltonians,
namely, in all four quartic cases, 
a general solution which is meromorphic and hyperelliptic with genus two.
As a consequence, no additional autonomous term can be added to 
either the cubic or the quartic Hamiltonians
without destroying the Painlev\'e integrability
(completeness property).
\end{abstract}


\noindent \textit{Keywords}:
H\'enon-Heiles Hamiltonian,
Painlev\'e property,
hyperelliptic.
separation of variables,


\noindent \textit{PACS 1995}:
02.30.Hq, 
03.40     

\baselineskip=14truept 




\section{Introduction}
\indent

The considered Hamiltonian originates from celestial mechanics,
as a system describing the motion of a star in the axisymmetric potential
of the galaxy.
Denoting $q_1$ the radius and $q_2$ the altitude,
this ``H\'enon-Heiles Hamiltonian'' (HH) \cite{HH}
is the sum of a kinetic energy and a potential energy,
in which the potential is a cubic polynomial in the position variables
$q_1,q_2$,
\begin{eqnarray}
H
& =&
 \frac{1}{2} (p_1^2 + p_2^2+ q_1^2 + q_2^2)+ q_1 q_2^2 - \frac{1}{3} q_1^3,\
\label{eqHHOriginal}
\end{eqnarray}
it is nonintegrable and displays a strange attractor.
However, if one changes the numerical coefficients in the potential,
the system may become integrable, 
and this question (to find all the integrable cases and to integrate them)
has attracted a lot of activity in the last three decennia.

A prerequisite is to define the word \textit{integrability},
and in section \ref{sectionIntegrability}
we briefly recall its three main acceptations in the context
of Hamiltonian systems.

In section \ref{sectionThe_seven},
we recall all the cases (three ``cubic'' plus four ``quartic'')
for which the most general two-degree of freedom classical 
time-independent Hamiltonian may have a single valued general solution.

Then,
discarding the integrated cases
(see \cite{CMVCalogero} for a review of the current state of this problem),
we focus on the three cases (all ``quartic'')
for which the general solution is still missing,
with the aim of finding this general solution.

In section \ref{sectionEquivalent_ODEs},
we build an equivalent fourth order 
ordinary differential equation (ODE) for $q_1(t)$,
in the hope of finding it listed in one of the classical tables
of explicitly integrated ODEs.
This hope is deceived because these tables are not yet finished.

This is why, in the last two sections,
we adopt a different strategy.
In front of the difficulty to perform the separation of variables
in the sense of Arnol'd and Liouville,
we establish a birational transformation between the two second order
Hamilton equations and a fourth order ODE listed in a classical table
established by Cosgrove \cite{Cos2000a},
whose general solution is single valued.

\section     {Integrability for Hamiltonian systems}
\label{sectionIntegrability}
\indent

Given a Hamiltonian system with a finite number $N$ of degrees of freedom,
three main definitions of \textit{integrability} are known,
\begin{enumerate}
\item
the one in the sense of Liouville,
that is the existence of $N$ independent invariants $K_j$
whose pairwise Poisson brackets vanish,
$\left\lbrace K_j,K_l \right\rbrace=0$,
\item
the one in the sense of Arnol'd-Liouville
\cite[chap.~9]{ArnoldMechanics},
\index{separating variables}
\index{Arnol'd-Liouville}
which is to find explicitly some canonical variables $s_j,r_j,j=1,N$
which ``separate'' the
\textit{Hamilton-Jacobi equation} for the action $S$,
 \index{Hamilton-Jacobi equation}
which for two degrees of freedom writes as,
\begin{eqnarray}
& &
H(q_1,q_2,p_1,p_2)-E=0,\
p_1=\pard{S}{q_1},\
p_2=\pard{S}{q_2},
\label{eqHJAutonomous}
\end{eqnarray}
\item
the one in the sense of Painlev\'e \cite{Cargese1996Conte}
i.e.~the representation of the general solution $q_j(t)$
by an explicit, closed form, single valued expression of the time $t$.
\end{enumerate}

\section     {The seven H\'enon-Heiles Hamiltonians}
\label{sectionThe_seven}
\indent
Given the most general
two-degree of freedom classical time-independent Hamiltonian
\begin{eqnarray}
H & =& \frac{1}{2} (p_1^2 + p_2^2) + V(q_1,q_2) =E,
\label{eqH2TV}
\end{eqnarray}
the requirement that the system made of the two Hamilton equations
passes the Painlev\'e test \cite{Cargese1996Conte}
(for at least some integer powers $q_1^{n_1}, q_2^{n_2}$)
selects seven and only seven potentials $V$
depending on a finite number of constants,
namely
\begin{enumerate}
\item
three ``cubic'' potentials (HH3 case) \cite{CTW,Fordy1991,CFP1993},
       \index{H\'enon-Heiles Hamiltonian!cubic}
\begin{eqnarray}
H
& =&
 \frac{1}{2} (p_1^2 + p_2^2 + \omega_1 q_1^2 + \omega_2 q_2^2)
    + \alpha q_1 q_2^2 - \frac{1}{3} \beta q_1^3
    + \frac{1}{2} \gamma q_2^{-2},\
\alpha \not=0
\label{eqHH0}
\end{eqnarray}
in which the constants $\alpha,\beta,\omega_1,\omega_2,\gamma$
can only take three sets of values,
\begin{eqnarray}
\hbox{(SK)} : & & \beta/ \alpha=-1, \omega_1=\omega_2,\
\label{eqHH3SKcond} \\
\hbox{(KdV5)} : & & \beta/ \alpha=-6,\
\label{eqHH3K5cond}\\
\hbox{(KK)} : & & \beta/ \alpha=-16, \omega_1=16 \omega_2.
\label{eqHH3KKcond}
\end{eqnarray}
\item
four ``quartic'' potentials (HH4 case) \cite{RDG1982,GDR1983},
\index{H\'enon-Heiles Hamiltonian!quartic}
\begin{eqnarray}
H & = &
\frac{1}{2}(P_1^2+P_2^2+\Omega_1 Q_1^2+\Omega_2 Q_2^2)
 +C Q_1^4+ B Q_1^2 Q_2^2 + A Q_2^4
\nonumber
\\
& &
 +\frac{1}{2}\left(\frac{\alpha}{Q_1^2}+\frac{\beta}{Q_2^2}\right)
 - \gamma Q_1,\ B \not=0,
\label{eqHH40}
\end{eqnarray}
in which the constants $A,B,C,\alpha,\beta,\gamma,\Omega_1,\Omega_2$
can only take the four values
(the notation $A:B:C=p:q:r$ stands for $A/p=B/q=C/r=\hbox{arbitrary}$),
\begin{eqnarray}
& & \left\lbrace
\begin{array}{ll}
\displaystyle{
A:B:C=1:2:1,\ \gamma=0,
}
\\
\displaystyle{
A:B:C=1:6:1,\ \gamma=0,\ \Omega_1=\Omega_2,
}
\\
\displaystyle{
A:B:C=1:6:8,\ \alpha=0,\ \Omega_1=4\Omega_2,
}
\\
\displaystyle{
A:B:C=1:12:16,\ \gamma=0,\ \Omega_1=4\Omega_2.
}
\end{array}
\right.
\label{eqHH4NLScond}
\end{eqnarray}
\end{enumerate}
All seven cases are integrable in the sense of Liouville,
with a second constant of the motion $K$
\cite{Drach1919KdV,BEF1995b,H1984} 
\cite{H1987,BakerThesis,BEF1995b} 
either quadratic or quartic in the momenta $p_1,p_2$.

In the sense of Arnol'd-Liouville,
the separation of variables has been performed
\cite{Drach1919KdV,Woj1985,RGC,VMC2002a,BSV1982,RRG},
except in three cases,
\begin{enumerate}
\item
HH4 1:6:1 $\alpha \not= \beta$,
\item
HH4 1:6:8 $\beta \gamma\not=0$,
\item
HH4 1:12:16 $\alpha \beta \not=0$.
\end{enumerate}

What is remarkable is the fact that,
in all cases when the separation of variables is achieved,
the equations of Hamilton have the Painlev\'e property,
the general solution being a hyperelliptic function of genus two.
The purpose of this work is to prove equally the Painlev\'e property
in the three remaining cases where the separation of variables is not yet
performed.

\section     {Equivalent fourth order ODEs}
\label{sectionEquivalent_ODEs}
\indent

In the cubic case,
the two Hamilton equations
\begin{eqnarray}
& & q_1'' + \omega_1 q_1 - \beta q_1^2 + \alpha q_2^2 = 0,
\label{eqHH1}
\\
& & q_2'' + \omega_2 q_2 + 2 \alpha q_1 q_2 - \gamma q_2^{-3}
=0,
\label{eqHH2}
\end{eqnarray}
together with the Hamiltonian (\ref{eqHH0}),
are equivalent \cite{Fordy1991}
to a single fourth order ODE for $q_1(t)$,
\begin{eqnarray}
& &
 q_1'''' + (8 \alpha - 2 \beta) q_1 q_1''
 - 2 (\alpha + \beta) q_1'^2
 - \frac{20}{3} \alpha \beta q_1^3
\nonumber
\\
& &
  +(\omega_1 + 4 \omega_2) q_1''
  + (6 \alpha \omega_1 - 4 \beta \omega_2) q_1^2 + 4 \omega_1 \omega_2 q_1
   + 4 \alpha E
=0,
\label{eqHH3ODE4}
\end{eqnarray}
independent of the coefficient $\gamma$ of the nonpolynomial term $q_2^{-2}$
and depending on the constant value $E$ of the Hamiltonian $H$.
In the three HH3 cases (\ref{eqHH3SKcond})--(\ref{eqHH3KKcond}),
this ODE belongs to a list \cite{Cos2000a} (``classification'')
of equations enjoying the Painlev\'e property,
whose general solution is hyperelliptic with genus two.

In the quartic case,
the similar fourth order equation 
is built by eliminating $Q_2$ and ${Q_1'''}^2$ between the 
two Hamilton equations,
\begin{eqnarray}
& & Q_1''+\Omega_1 Q_1 + 4 C Q_1^3 + 2 B Q_1 Q_2^2 - \alpha Q_1^{-3} + \gamma=0,
\label{eqHH41}
\\
& & Q_2''+\Omega_2 Q_2 + 4 A Q_2^3 + 2 B Q_2 Q_1^2 - \beta  Q_2^{-3}=0,
\label{eqHH42}
\end{eqnarray}
and the Hamiltonian (\ref{eqHH40}), which results in
\begin{eqnarray}
& &
-Q_1''''
+ 2 \frac{Q_1' Q_1'''}{Q_1}
+\left(1 + 6 \frac{A}{B}\right) \frac{{Q_1''}^2}{Q_1}
-2 \frac{{Q_1'}^2 Q_1''}{Q_1^2}
\nonumber \\ & & \phantom{12345}
+8 \left(6 \frac{A C}{B} - B - C\right) Q_1^2 Q_1''
+4(B - 2 C) Q_1 {Q_1'}^2
+24 C \left(4 \frac{A C}{B} - B\right) Q_1^5
\nonumber \\ & & \phantom{12345}
+\left\lbrack
12 \frac{A}{B} \omega_1 - 4 \omega_2
+\left(1 + 12 \frac{A}{B}\right) \frac{\gamma}{Q_1}
- 4 \left(1+3 \frac{A}{B}\right) \frac{\alpha}{Q_1^4}
\right\rbrack Q_1''
\nonumber \\ & & \phantom{12345}
+ 6 \frac{A}{B} \frac{\alpha^2}{Q_1^7}
+ 20 \frac{\alpha} {Q_1^5} {Q_1'}^2
-12  \frac{A}{B} \frac{\gamma \alpha}{Q_1^4}
+4 \left(3 \frac{A}{B} \omega_1 - \omega_2 \right)
   \left(\gamma - \frac{\alpha}{Q_1^3}\right)
-2 \gamma \frac{{Q_1'}^2}{Q_1^2}
\nonumber \\ & & \phantom{12345}
+ 6 \left(\frac{A}{B} \gamma^2 + 2 B \alpha -8 \frac{A C}{B} \alpha\right)
     \frac{1}{Q_1}
+ \left(6 \frac{A}{B} \omega_1^2 -4 \omega_1 \omega_2 -8 B E\right) Q_1
\nonumber \\ & & \phantom{12345}
+ 48 \frac{A C}{B} \gamma Q_1^2
+ 4 \left(12  \frac{A C}{B} - B - 4 C \right) \omega_1 Q_1^3.
\label{eqHH4odeq1}
\end{eqnarray}
This ODE depends on $E$ but not on $\beta$
and, as opposed to the cubic case,
it does not belong to a classified set of equations,
because $Q_1''''$ is not polynomial in $Q_1$.

In the three remaining cases,
since one is yet unable 
either to perform the separation of variables
or to establish a direct link to a classified ODE,
let us build an indirect link to such a classified ODE.
This link, which involves soliton equations, is the following.

For each of the seven cases,
the two Hamilton equations are equivalent
\cite{Fordy1991,FK1983,BakerThesis} 
to the traveling wave reduction of a soliton system made 
either of a single PDE (HH3) or of two coupled PDEs (HH4),
most of them appearing in lists established from group theory
\cite{DS1981}.
Among the various soliton equations which are equivalent to them
\textit{via} a B\"acklund transformation,
some of them admit a traveling wave reduction to a classified ODE.
This property defines a path \cite{MV2003,VMC2004a} which
starts from one of the three remaining HH4 cases,
goes up to a soliton system of two coupled 1+1-dimensional PDEs
admitting a reduction to the considered case,
then goes to another 1+1-dim PDE system equivalent under a
B\"acklund transformation,
finally goes down by reduction to an already integrated
ODE or system of ODEs.

\section     {General solution of the quartic 1:6:1 and 1:6:8 cases}
\label{sectionIntegrationQuartic1:6:1and1:6:8}
\indent

Let us denote the two constants of the motion of the 1:6:1 and 1:6:8 cases
as,
\begin{eqnarray}
{\hskip -15.0 truemm}
& &
1:6:1
\left\lbrace
\begin{array}{ll}
\displaystyle{
H =
 \frac{1}{2}(P_1^2+P_2^2)
+\frac{\Omega}{2}(Q_1^2+Q_2^2)
-\frac{1}{32} (Q_1^4+ 6 Q_1^2 Q_2^2 + Q_2^4)
}
\\
\displaystyle{
\phantom{1234}
 -\frac{1}{2}\left(\frac{\kappa_1^2}{Q_1^2}+\frac{\kappa_2^2}{Q_2^2}\right)
=E,
}
\\
\displaystyle{
K =
\left(
P_1 P_2 + Q_1 Q_2 \left(-\frac{Q_1^2+Q_2^2}{8}+\Omega \right)
\right)^2
}
\\
\displaystyle{
\phantom{1234}
- P_2^2 \frac{\kappa_1^2}{Q_1^2}
- P_1^2 \frac{\kappa_2^2}{Q_2^2}
+\frac{1}{4}\left(\kappa_1^2 Q_2^2 + \kappa_2^2 Q_1^2 \right)
+\frac{\kappa_1^2 \kappa_2^2}{Q_1^2 Q_2^2},
}
\end{array}
\right.
 \label{eqHH40161}
\end{eqnarray}
and
\begin{eqnarray}
{\hskip -15.0 truemm}
& &
1:6:8
\left\lbrace
\begin{array}{ll}
\displaystyle{
H =
 \frac{1}{2}(p_1^2+p_2^2)
+\frac{\omega}{2}(4 q_1^2+q_2^2)
-\frac{1}{16} (8 q_1^4+ 6 q_1^2 q_2^2 + q_2^4)
}
\\
\displaystyle{
\phantom{1234}
- \gamma q_1 +\frac{\beta}{2 q_2^2}
=E,
}
\\
\displaystyle{
K =
\left(
p_2^2-\frac{q_2^2}{16}(2 q_2^2+4 q_1^2+\omega)
     +\frac{\beta}{q_2^2}
\right)^2
-\frac{1}{4}q_2^2(q_2 p_1 - 2 q_1 p_2)^2
}
\\
\displaystyle{
\phantom{1234}
+\gamma
\left(
-2 \gamma q_2^2
-4 q_2 p_1 p_2
+\frac{1}{2} q_1 q_2^4
+ q_1^3 q_2^2
+4 q_1 p_2^2
-4 \omega q_1 q_2^2
+ 4 q_1 \frac{\beta}{q_2^2}
\right).
}
\end{array}
\right.
\label{eqHH40168}
\end{eqnarray}
There is a canonical transformation \cite{BakerThesis}
between the 1:6:1 and 1:6:8 cases, mapping the constants as follows,
\begin{eqnarray}
E_{1:6:8}=E_{1:6:1},\
K_{1:6:8}=K_{1:6:1},\
\omega=\Omega,\
\gamma=\frac{\kappa_1+\kappa_2}{2},\
\beta=- (\kappa_1-\kappa_2)^2,
\label{eqCT161to168constants}
\end{eqnarray}
therefore one only needs to integrate either case.

The path to an integrated ODE comprises the following three segments.

The coordinate $q_1(t)$ of the 1:6:8 case can be identified
\cite{BEF1995b,BakerThesis}
to the component $F$ of
the traveling wave reduction
$f(x,\tau)=F(x-c \tau), g(x,\tau)=G(x-c \tau)$
of a soliton system of two coupled KdV-like equations
(\cKdV\ system)
denoted \cKdV${}_1$
\cite{BEF1995b,BakerThesis}
\begin{eqnarray}
{\hskip -10.0 truemm}
& &
\left\lbrace
\begin{array}{ll}
\displaystyle{
f_\tau+ \left(f_{xx}+\frac{3}{2} f f_{x}-\frac{1}{2}f^3+3 f g\right)_x=0,
}
\\
\displaystyle{
- 2 g_\tau+ g_{xxx}+ 6 g g_x+3 f g_{xx}+6 gf_{xx}+9 f_x g_x-3 f^2 g_x
}
\\
\displaystyle{
\phantom{xxxxx}+\frac{3}{2} f_{xxxx} +\frac{3}{2} f f_{xxx}+9 f_x f_{xx}
               -3 f^2 f_{xx}-3 f f_x^2=0,
}
\end{array}
\right.
\label{eq:cKdV1}
\end{eqnarray}
with the identification
\begin{eqnarray}
{\hskip -10.0 truemm}
& &
\left\lbrace
\begin{array}{ll}
\displaystyle{
q_1=F,\
q_2^2=-2\left(F'+F^2 + 2 G -2 \omega\right),\
}
\\
\displaystyle{
c=-\omega,\
K_1=\gamma,\
K_2=E,
}
\end{array}
\right.
\label{eqcKdV1r_to_HH4168}
\end{eqnarray}
in which $K_1$ and $K_2$ are two constants of integration.

There exists a B\"acklund transformation between this soliton system
and another one of the \cKdV\ type, denoted bi-SH system \cite{DS1981},
\begin{eqnarray}
& &
\left\lbrace
\begin{array}{ll}
\displaystyle{
- 2 u_\tau+ \left(u_{xx} + u^2 + 6 v\right)_x = 0,
}
\\
\displaystyle{
v_\tau+ v_{xxx} + u v_x = 0.
}
\end{array}
\right.
\label{eq:systemHSII}
\end{eqnarray}
This BT is defined by the Miura transformation \cite{MV2003}
\begin{eqnarray}
{\hskip -10.0 truemm}
& &
\left\lbrace
\begin{array}{ll}
\displaystyle{
u=\frac{3}{2} \left(2 g -f_x-f^2\right),
}
\\
\displaystyle{
v=\frac{3}{4}
\left(
 2 f_{xxx}
+4 f f_{xx}
+8 g f_x
+4 f g_x
+3 f_x^2
-2 f^2 f_x
-  f^4
+4 g f^2\right).
}
\end{array}
\right.
\end{eqnarray}

Finally,
the traveling wave reduction 
$u(x,\tau)=U(x-c \tau),v(x,\tau)=V(x-c \tau)$
can be identified \cite{VMC2004a}
to the autonomous F-VI equation (a-F-VI)
in the classification of Cosgrove \cite{Cos2000a},
\begin{eqnarray}
& &
\hbox{a-F-VI}:\
y''''=18 y y'' + 9 {y'}^2 - 24 y^3
+ \alpha_{\rm VI} y^2 + \frac{\alpha_{\rm VI}^2}{9} y
+ \kappa_{\rm VI} t + \beta_{\rm VI},\
\kappa_{\rm VI}=0,
\label{eqCosgroveFVI}
\end{eqnarray}
an ODE whose general solution is meromorphic,
expressed with genus two hyperelliptic functions \cite[Eq.~(7.26)]{Cos2000a}.
The identification is
\begin{eqnarray}
{\hskip -10.0 truemm}
& &
\left\lbrace
\begin{array}{ll}
\displaystyle{
U=- 6 \left(y+\frac{c}{18}\right),\
}
\\
\displaystyle{
V=y'' -6 y^2 + \frac{4}{3} c y
+\frac{16}{27} c^2 -\frac{K_A}{2},
}
\\
\displaystyle{
\alpha_{\rm VI}=-4 c,\
\beta_{\rm VI}=K_B-2 c K_A + \frac{512}{243} c^3,
}
\end{array}
\right.
\label{eqFVI_to_biSHr}
\end{eqnarray}
in which $K_A,K_B$ are two constants of integration.

In order to perform the integration of both the 1:6:1 and the 1:6:8 cases,
it is sufficient 
to express $(F,G)$ rationally in terms of $(U,V,U',V')$.
The result is
\begin{eqnarray}
{\hskip -10.0 truemm}
& &
\left\lbrace
\begin{array}{ll}
\displaystyle{
F=\frac{W'}{2 W}
+ \frac{K_1}{24 W}
\left[
-3 {U'}^2
-2 (U-3 c) \left(12 V + (U + 3 c)^2\right)
\right.
}
\\
{
\left.
\phantom{1234567890123456}
+36 K_B
-54 K_1^2
\right],
}
\\
\displaystyle{
G=\frac{U}{3} + \frac{1}{8 W}
\left[
(2 V + 3 K_2) \left(2 V'' + K_1 U' -3 K_1^2 \right)
\right.
}
\\
{
\left.
\phantom{1234567890123}
-2 (U - 3 c)\left(2 K_1 V' + K_1^2 (U + 3 c)\right)
\right],
}
\\
\displaystyle{
W=\left(V + \frac{3}{2} K_2\right)^2 + \frac{3}{2} K_1^2 (U - 3 c),
}
\\
\displaystyle{
K_A=K_2.
}
\end{array}
\right.
\label{eqbiSHr_to_cKdV1r}
\end{eqnarray}

Making the product of the successive transformations
(\ref{eqcKdV1r_to_HH4168}),
(\ref{eqbiSHr_to_cKdV1r}),
(\ref{eqFVI_to_biSHr}),
one obtains 
a meromorphic general solution
for $Q_1^2,Q_2^2,q_1,q_2^2$,
\begin{eqnarray}
{\hskip -10.0 truemm}
& &
\left\lbrace
\begin{array}{ll}
\displaystyle{
q_1=\frac{W'}{2 W}
+ \frac{\gamma}{W}
\left[9 j -3 \left(y + \frac{4}{9}\omega\right) (h+E)
      - \frac{9}{4} \gamma^2\right],
}
\\
\displaystyle{
q_2^2=-16 \left(y - \frac{5}{9} \omega\right)
}
\\
\displaystyle{
\phantom{123}
+\frac{1}{W}
\Big[
 12 \left(y' + \frac{\gamma}{2}\right)^2
 -48 y^3 - 16 \omega y^2
 +\left(24 E +\frac{128}{9}\omega^2\right) y
 + \frac{1280}{243} \omega^3
\Big.
} \\ \displaystyle{\phantom{12345678}
\Big.
- \frac{40}{3} \omega E + \frac{3}{4} \beta
-24 \gamma \left(y - \frac{5}{9} \omega\right) h'
-144 \gamma^2 \left(y - \frac{5}{9} \omega\right)^2
\Big],
}
\\
\displaystyle{
W= (h+E)^2 -9 \gamma^2 \left(y - \frac{5}{9} \omega\right),
}
\\
\displaystyle{
\alpha_{\rm VI}=4 \omega,\
\beta_{\rm VI}=\frac{3}{4} \gamma^2 + 2 \omega E
               - \frac{3}{16} \beta-\frac{512}{243} \omega^3,
}
\\
\displaystyle{
K_{1,{\rm VI}}=\frac{3}{32}   K - \frac{1}{2} E^2,\
K_{2,{\rm VI}}=\frac{3}{32} E K - \frac{1}{3} E^3+\frac{9}{64} \beta \gamma^2,
}
\\
\displaystyle{
K_1=\gamma,\
K_2=E,\
K_A=E,\
K_B=- \frac{3}{16} \beta + \frac{3}{4} \gamma^2.
}
\end{array}
\right.
\label{eqFVI_to_HH4168}
\end{eqnarray}
in which $h$ and $j$ are convenient auxiliary variables
\cite[Eqs.~(7.4)--(7.5)]{Cos2000a},
\begin{eqnarray}
{\hskip -10.0 truemm}
& &
\left\lbrace
\begin{array}{ll}
\displaystyle{
y=\frac{Q(s_1,s_2) + \sqrt{Q(s_1) Q(s_2)}}
       {2\left(\sqrt{s_1^2-C_{\rm VI}}+\sqrt{s_2^2-C_{\rm VI}} \right)^2}
 + \frac{5}{36} \alpha_{\rm VI},\
}
\\
\displaystyle{
h=-\frac{3}{4} E_{\rm VI}
 \frac{s_1 s_2 + C_{\rm VI} + \sqrt{(s_1^2-C_{\rm VI})(s_2^2-C_{\rm VI})}}
      {s_1+s_2} 
 - \frac{F_{\rm VI}}{2},\
}
\\
\displaystyle{
j=\frac{1}{6} (2 h + F_{\rm VI})
 \left\lbrace
y + \frac{\alpha_{\rm VI}}{9} - \frac{E_{\rm VI}}{4(s_1+s_2)}.
\right\rbrace
}
\end{array}
\right.
\label{eqFVI_General_solution}
\end{eqnarray}
In the above,
the variables $s_1,s_2$ are defined by the
\textit{hyperelliptic system} \cite{Cos2000a}
\begin{eqnarray}
& &
\left\lbrace
\begin{array}{ll}
\displaystyle{
(s_1-s_2)s_1'=\sqrt{P(s_1)},\
(s_2-s_1)s_2'=\sqrt{P(s_2)},\
}
\\
\displaystyle{
P(s)=(s^2-C_{\rm VI}) Q(s),
}
\\
\displaystyle{
Q(s,t)=(s^2-C_{\rm VI})(t^2-C_{\rm VI}) 
- \frac{\alpha_{\rm VI}}{2} (s^2+t^2-2 C_{\rm VI})
        +\frac{E_{\rm VI}}{2}(s+t)+F_{\rm VI},
}
\\
\displaystyle{
Q(s)=Q(s,s).
}
\end{array}
\right.
\label{eqHyperellipticGenusTwoSystem}
\label{eqCosgroveFVIsextic}
\end{eqnarray}
The expressions (\ref{eqFVI_General_solution})
cannot be written as rational functions of $s_1,s_2,s_1',s_2'$
and are nevertheless meromorphic \cite{FarkasKra,Mumford}.

The coefficients $(\alpha,C,E,F)_{\rm VI}$
of the hyperelliptic curve 
depend algebraically 
on the parameters of the Hamiltonians
$\beta,\gamma,\omega,E,K$ 
\cite[Eqs.~(7.9)-(7.12)]{Cos2000a}
\begin{eqnarray}
{\hskip -10.0 truemm}
& &
\left\lbrace
\begin{array}{ll}
\displaystyle{
A_{\rm VI}=4 \omega,
}
\\
\displaystyle{
E_{\rm VI}^2=-\frac{16}{3}\omega(F_{\rm VI}-2 E)-\beta + 4 \gamma^2,
}
\\
\displaystyle{
C_{\rm VI} E_{\rm VI}^2=\frac{4}{3}(F_{\rm VI}^2-4 E^2) +K,
}
\\
\displaystyle{
(F_{\rm VI}-2 E)^2 (F_{\rm VI}+4 E) + \frac{9 K}{4} (F_{\rm VI}-2 E)
- \frac{27}{4} \beta \gamma^2=0,
}
\end{array}
\right.
\label{eqFVI_to_HH4168_const}
\end{eqnarray}
and this algebraic dependence could explain the difficulty 
to separate the variables in the Hamilton-Jacobi equation.
Note that,
in the particular case $\beta \gamma=0$, i.e.~$\kappa_1^2=\kappa_2^2$,
these coefficients become rational, see \cite{VMC2003}.

\textit{Remark}.
The F-VI ODE can be written in Hamiltonian form,
\begin{eqnarray}
& & \left\lbrace
\begin{array}{ll}
\displaystyle{
H=P_2^2 
+ Q_2 P_1
-\frac{Q_1^4}{6}
+\frac{3}{2} Q_1 Q_2^2
-\frac{13}{1296} \alpha_{\rm VI}^3 Q_1
+\frac{1}{16} \alpha_{\rm VI}^2 Q_1^2
-\frac{1}{8} \alpha_{\rm VI} Q_2^2
}
\\
\displaystyle{
\phantom{1234}
- 6 \beta_{\rm VI} Q_1
- 6 \kappa_{\rm VI} t Q_1
+\frac{347}{2^9 3^3}  \alpha_{\rm VI}^4
+\frac{9}{2} \alpha_{\rm VI} \beta_{\rm VI},
}
\\
\displaystyle{
Q_1=-6 \left(y-\frac{\alpha_{\rm VI}}{72} \right),\ 
Q_2=-6 y',\
P_1= 6 y''' - 108 y y',\
P_2=-6 y''.
}
\end{array}
\right.
\label{eqHbF-VI_Ham}
\end{eqnarray}
In the autonomous case $\kappa_{\rm VI}=0$,
the Hamiltonian $H$ is a first integral 
(equal to $36 \KFVIone$),
and the other constant of the motion is cubic in the momenta.
However,
because of the nonlinear link between $\KFVIone$ and the two first integrals
of the 1:6:8 case,
see (\ref{eqFVI_to_biSHr}),
there exists no canonical transformation between
the variables $(q_j,p_j)$ of 1:6:8 
and the above canonical variables of a-F-VI.

\section     {General solution of the quartic 1:12:16 case}
\label{sectionIntegrationQuartic1:12:16}
\indent
Let us denote the two constants in involution as,
\begin{eqnarray}
& &
1:12:16
\left\lbrace
\begin{array}{ll}
\displaystyle{
H =
\frac{1}{2}(P_1^2+P_2^2) + \frac{\Omega}{8} (4 Q_1^2+ Q_2^2)
 - \frac{1}{32} (16 Q_1^4+ 12 Q_1^2 Q_2^2 + Q_2^4)
}
\\
\displaystyle{
\phantom{1234}
 -\frac{1}{2}\left(\frac{\kappa_1^2}{Q_1^2}+\frac{4 \kappa_2^2}{Q_2^2}\right)
=E,
}
\\
\displaystyle{
K =
\frac{1}{16}
\left(8 (Q_2 P_1 - Q_1 P_2) P_2 - Q_1 Q_2^4 - 2 Q_1^3 Q_2^2
 + 2 \Omega Q_1 Q_2^2 + 32 Q_1 \frac{\kappa_2^2}{Q_2^2} \right)^2
}
\\
\displaystyle{
\phantom{1234}
+ \kappa_1^2 \left(Q_2^4 - 4 \frac{Q_2^2 P_2^2}{Q_1^2}\right).
}
\end{array}
\right.
\label{eqHH401:12:16E}
\label{eqHH401:12:16K}
\end{eqnarray}

Similarly to the 1:6:1-1:6:8 couple,
there exists a canonical transformation between the 1:12:16 Hamiltonian
and another Hamiltonian \cite{BakerThesis,BEF1995b},
which is however not the sum of a kinetic energy and a potential energy,
which we denote similarly as 5:9:4,
\begin{eqnarray}
& &
5:9:4
\left\lbrace
\begin{array}{ll}
\displaystyle{
H =
\frac{1}{2}\left(p_1^2+\left(p_2 - \frac{3}{2} q_1 q_2\right)^2\right)
-\frac{1}{8} (4 q_1^4 + 9 q_1^2 q_2^2 +5 q_2^4)
+\frac{\omega}{2}(q_1^2+q_2^2)
-\kappa q_1
+\frac{\zeta}{2 q_2^2}
=E,
}
\\
\displaystyle{
K=\frac{1}{q_2^2}
\left(
  2 q_2^2 p_1 
+ 2 q_1^2 q_2^2
-2 q_1 q_2 p_2
- q_2^4
 - 4 \kappa q_1
\right)^2
}
\\
\displaystyle{
\phantom{12345}
\times
\left(
  2 q_2^2 p_1 
+ 2 q_1^2 q_2^2
+ p_2^2
-4 q_1 q_2 p_2
- 2 q_2^4
 + \omega q_2^2
 + 4\frac{\kappa^2}{q_2^2}
 + 8 \kappa q_1
 -4 \kappa \frac{p_2}{q_2}
\right)
}
\\
\displaystyle{
\phantom{1234}
+ 4 (\zeta + 4 \kappa^2)
\left(
\left(-2 q_1 \frac{p_2}{q_2}+4 q_1^2+q_2^2+4 q_1 \frac{\kappa}{q_2^2}\right) p_1
 - \frac{1}{q_2^4} (q_1^2 q_2^2 + q_2^4 + 2 \kappa q_1)^2
\right.
}
\\
\displaystyle{
\phantom{123456789012345}
\left.
+2 \frac{q_1^2}{q_2^2} \left(p_2 - \frac{3}{2} q_1 q_2 \right)^2
+ \frac{(q_1^2+q_2^2)^2}{2}
+q_1^2 \frac{\zeta}{q_2^4}
\right),
}
\\
\displaystyle{
E_{5:9:4}=E_{1:12:16},\
K_{5:9:4}=K_{1:12:16},\
\omega=\Omega,\
\kappa=\frac{\kappa_1 + \kappa_2}{2},\
\zeta=-(\kappa_1 - \kappa_2)^2.
}
\end{array}
\right.
\label{eqHH405:9:4E}
\label{eqHH405:9:4K}
\end{eqnarray}

The path to an integrated ODE is also quite similar and is made
of the following three segments \cite{BakerThesis,V2003,MV2003}.

Firstly,
the coordinate $q_1(t)$ of (\ref{eqHH405:9:4E}) is identified
\cite{BEF1995b,BakerThesis}
to the component $F$ of the traveling wave reduction
$f(x,\tau)=F(x-c \tau), g(x,\tau)=G(x-c \tau)$
of a soliton system of two coupled KdV-like equations
denoted {\cKdV}a$(f,g)$
\cite{BEF1995b,BakerThesis},
\begin{eqnarray}
{\hskip -10.0 truemm}
& &
\left\lbrace
\begin{array}{ll}
\displaystyle{
q_1=F,\
q_2^2=\frac{2}{5}\left(F'-2 F^2 -G + \omega\right),
}
\\
\displaystyle{
c=-\omega.
}
\end{array}
\right.
\end{eqnarray}

Secondly,
there exists a B\"acklund transformation between this soliton system
and another one of the \cKdV\ type, 
denoted bi-SK system $(u,v)$ \cite{Ramani1981},
transformation defined by the Miura map
\begin{eqnarray}
& &
\left\lbrace
\begin{array}{ll}
\displaystyle{
u=\frac{3}{10} \left(3 f_x - f^2 + 2 g \right),
}
\\
\displaystyle{
v=\frac{9}{10}\left(f_{xxx} + g_{xx} + f_x g - f g_x -f f_{xx} + g^2\right).
}
\end{array}
\right.
\label{eqMiura_from_cKdVa_to_biSK}
\end{eqnarray}

Finally,
the traveling wave reduction 
$u(x,\tau)=U(x-c \tau),v(x,\tau)=V(x-c \tau)$
is identified \cite{VMC2004a},
\begin{eqnarray}
{\hskip -10.0 truemm}
& &
\begin{array}{ll}
\displaystyle{
U=-3 \left(y - \frac{\omega}{30}\right),\
V=-6 y'' +18 y^2 - \frac{9}{5} \omega y +\frac{1}{10} \omega^2-\frac{3}{5} E,
}
\end{array}
\end{eqnarray}
to the F-IV equation (or to the F-III as well)
in the classification of Cosgrove \cite{Cos2000a},
\begin{eqnarray}
& \hbox{F-IV} & 
\left\lbrace
\begin{array}{ll}
\displaystyle{
y''''=30 y y'' - 60 y^3 + \alpha_{\rm IV} y + \beta_{\rm IV},
}
\\
\displaystyle{
y=\frac{1}{2}
 \left(s_1'+s_2' + s_1^2+s_1 s_2 + s_2^2 +A \right),
}
\\
\displaystyle{
(s_1-s_2)s_1'=\sqrt{P(s_1)},\
(s_2-s_1)s_2'=\sqrt{P(s_2)},\
}
\\
\displaystyle{
P(s)=(s^2+A)^3-\frac{\alpha_{\rm IV}}{3} (s^2+A) + B s + \frac{\beta_{\rm IV}}{3},
}
\\
\displaystyle{
\KFIVone=\left(\frac{3 B}{4} \right)^2,\
\KFIVtwo=-\frac{9 A B^2}{64},
}
\end{array}
\right.
\label{eqCosgroveFIV}
\end{eqnarray}
in which $(\KFIVone,\KFIVtwo)$ denote two polynomial first integrals of F-IV.
The general solution of this ODE is meromorphic,
expressed with genus two hyperelliptic functions \cite{Cos2000a}.

In order to perform the integration of both Hamiltonians
(\ref{eqHH401:12:16E}) and (\ref{eqHH405:9:4E}),
it is sufficient 
to express $(F,G)$ rationally in terms of $(U,V,U',V')$.
The result is
\begin{eqnarray}
{\hskip -10.0 truemm}
& &
\left\lbrace
\begin{array}{ll}
\displaystyle{
F=-\frac{W'}{2 W} + \Kaone X_2,
}
\\
\displaystyle{
G=-F^2 - X_1 X_2 + \Kaone \frac{54 U'}{X_1}
 - 54 \Kaone  \left(U + \frac{3 \omega}{20}\right) \frac{W'}{W X_1}
 + \frac{2}{3}\left(U + \frac{9 \omega}{10}\right),
}
\\
\displaystyle{
W=X_1^2 +108 \Kaone^2 \left(U + \frac{3 \omega}{20}\right),\
}
\\
\displaystyle{
X_1=V + 2 U^2 - 3 \omega U + \frac{9}{50} \omega^2 - \frac{27}{5} E,
}
\\
\displaystyle{
X_2=
 9 \left(
-4 {U'}^2
+ \frac{8}{3} U V
- \frac{8}{25} \omega U^2
+ \frac{2}{5} \omega V
+ \frac{48}{5} E U
\right.
}
\\
\displaystyle{
\left.
\phantom{12345xxxxxxxxxx}
- \frac{42}{25} \omega^2 U
- \frac{9}{2} (\kappa_1^2 + \kappa_2^2)
- \frac{9}{2} \Kaone^2
+ \frac{36}{25} \omega E
- \frac{27}{125} \omega^3
\right),
}
\\
\displaystyle{
\Kaone=\kappa_1 - \kappa_2.
}
\end{array}
\right.
\label{eqbira_from_cKdVa_to_biSK}
\end{eqnarray}

{}From the point of view of the separation of variables,
one should first exhibit a Hamiltonian representation of F-IV.
One such structure is that of the cubic SK case.
However,
since the constant value of the Hamiltonian of the cubic SK case,
when expressed only in terms of the parameters
$(E,K,\omega,\kappa_1,\kappa_2)$ of the 1:12:16,
is not an affine function of $E$,
there exists no canonical transformation between
the cubic SK case and the 1:12:16 case.

\section{Conclusion, remaining work}
\label{sectionOpen_problems}

The explicit integration of all the seven cases
is now achieved in the Painlev\'e sense
(finding a closed form single valued expression for the general solution),
and the common features are the following.
\begin{enumerate}
\item
In all cases, the general solution is hyperelliptic with genus two,
and therefore meromorphic.
\item
Each case is birationally equivalent to a fourth order ODE
which is \textit{complete} in the Painlev\'e sense,
i.e.~which accepts no additional term,
under penalty of losing its Painlev\'e property.
Consequently,
for each of the seven Hamiltonians,
it is impossible to add any term to the Hamiltonian without destroying
the Painlev\'e property,
and the seven H\'enon-Heiles Hamiltonians are complete.
\end{enumerate}

About the integration in the Arnol'd-Liouville sense
(finding the separating variables of the Hamilton-Jacobi equation),
two problems remain open.
\begin{enumerate}
\item
In the 1:6:1-1:6:8 case,
the hyperelliptic curve $y^2=P(s)$ of F-VI (see (\ref{eqCosgroveFVIsextic}))
reduces in the separated cases $\beta \gamma=0$
to the hyperelliptic curve of the separating variables.
Therefore, F-VI is the good ODE to consider,
and the only missing item is to find a Hamiltonian structure of F-VI,
necessarily distinct from (\ref{eqHbF-VI_Ham}),
admitting a canonical transformation to 1:6:1-1:6:8.

\item
In the 1:12:16-5:9:4 case,
the hyperelliptic curve $y^2=P(s)$ of F-IV (see (\ref{eqCosgroveFIV}))
does not reduce in the separated cases $\kappa_1 \kappa_2=0$ \cite{V2003}
to the hyperelliptic curve of the separating variables, which is
\begin{eqnarray}
& &
\kappa_1 \kappa_2=0:\
P(s)=s^6 - \omega s^3 + 2 E s^2 + \frac{K}{20} s
 + \kappa_1^2 + \kappa_2^2=0.
\label{eq1:12:16kappa1kappa20}
\end{eqnarray}
Therefore, F-IV (as well as its birationally equivalent ODE F-III)
is not the good ODE to consider,
and it should be quite instructive to integrate the fourth order 
equivalent ODE (\ref{eqHH4odeq1}) in that case.
\end{enumerate}

\section*{Acknowledgments}

The authors acknowledge the financial support of
the Tournesol grant no.~T2003.09.
CV is a postdoctoral fellow at the FWO-Vlaanderen.


\label{lastpage}
\vfill \eject

\begin{thebibliography}{99}
\small

\bibitem{AF2000} S.~Abenda and Yu.~Fedorov,
Acta Appl.~Math.~{\bf 60} (2000) 137--178.

\bibitem{ArnoldMechanics} V.I.~Arnol'd,
\textit{Les m\'ethodes math\'ematiques de la m\'ecanique classique}
(Nauka, Moscou, 1974)
(Mir, Moscou, 1976).

\bibitem{BakerThesis} S.~Baker,
PhD Thesis, University of Leeds (1995).

\bibitem{BEF1995b} S.~Baker, V.~Z.~Enol'skii, and A.~P.~Fordy,
Phys.~Lett.~A {\bf 201} (1995) 167--174.

\bibitem{BW1994} M.~B{\l}aszak and S.~Rauch-Wojciechowski,
J.~Math.~Phys.~{\bf 35} (1994) 1693--1709.

\bibitem{BSV1982} T.~Bountis, H.~Segur, and F.~Vivaldi,
Phys.~Rev.~A {\bf 25} (1982) 1257--1264.

\bibitem{CTW} Chang Y.~F., M.~Tabor, and J.~Weiss,
J.~Math.~Phys.~{\bf 23} (1982) 531--538.

\bibitem{Cargese1996Conte} R.~Conte,
{\it The Painlev\'e property, one century later},
77--180,
ed.~R.~Conte,
CRM series in mathematical physics (Springer, New York, 1999).
solv-int/9710020.

\bibitem{CFP1993} R.~Conte, A.~P.~Fordy, and A.~Pickering,
Physica D {\bf 69} (1993) 33--58.

\bibitem{CMVCalogero} R.~Conte, M.~Musette, and C.~Verhoeven,
J.~Nonlinear Math.~Phys.~{\bf 12} Supp.~1 (2005) 212--227.
nlin.SI/0412057.

\bibitem{Cos2000a} C.~M.~Cosgrove,
Stud.~Appl.~Math.~{\bf 104} (2000) 1--65.

\bibitem{Drach1919KdV} J.~Drach,
\CRAS\ {\bf 168} (1919) 337--340.

\bibitem{DS1981} V.~G.~Drinfel'd and V.~V.~Sokolov,
Soviet Math.~Dokl.~{\bf 23} (1981) 457--462.

\bibitem{FarkasKra} N.~Farkas and E.~Kra,
\textit{Riemann surfaces} (Springer, Berlin, 1980).

\bibitem{Fordy1991} A.~P.~Fordy,
Physica D {\bf 52} (1991) 204--210.

\bibitem{FK1983} A.P.~Fordy and P.~P.~Kulish,
Commun.~Math.~Phys.~{\bf 89} (1983) 427--443.

\bibitem{GDR1983} B.~Grammaticos, B.~Dorizzi, and A.~Ramani,
J.~Math.~Phys.~{\bf 24} (1983) 2289--2295.

\bibitem{HH} M.~H\'enon and C.~Heiles,
Astron.~J.~{\bf 69} (1964) 73--79.

\bibitem{H1984} J.~Hietarinta,
J.~Math.~Phys.~{\bf 25} (1984) 1833--1840.

\bibitem{H1987} J.~Hietarinta,
Phys.~Rep.~{\bf 147} (1987) 87--154.

\bibitem{Mumford} D.~Mumford,
\textit{Tata lectures on theta, II} (Birkh\"auser, Basel, 1983).

\bibitem{MV2003} M.~Musette and C.~Verhoeven,
Theor.~Math.~Phys.~{\bf 137} (2003) 1561--1573.

\bibitem{Ramani1981} A.~Ramani,                      
in {\it Fourth international conference on collective phenomena},
ed.~J.~L.~Lebowitz,
Annals of the New York Academy of Sciences {\bf 373} 54--67
(NY Ac.~Sc., New York, 1981).

\bibitem{RDG1982} A.~Ramani, B.~Dorizzi, and B.~Grammaticos,
Phys.~Rev.~Lett.~{\bf 49} (1982) 1539--1541.

\bibitem{RGC} V.~Ravoson, L.~Gavrilov, and R.~Caboz,
J.~Math.~Phys.~{\bf 34} (1993) 2385--2393.

\bibitem{RRG} V.~Ravoson, A.~Ramani, and B.~Grammaticos,
Phys.~Letters A {\bf 191} (1994) 91--95.

\bibitem{V2003} C.~Verhoeven,
PhD thesis, Vrije Universiteit Brussel (May 2003).

\bibitem{VMC2002a} C.~Verhoeven, M.~Musette, and R.~Conte,
J.~Math.~Phys.~{\bf 43} (2002) 1906--1915.  nlin.Si/0112030.

\bibitem{VMC2003} C.~Verhoeven, M.~Musette, and R.~Conte,
Theor.~Math.~Phys.~{\bf 134}, 128--138 (2003).
nlin.SI/0301011.

\bibitem{VMC2004a} C.~Verhoeven, M.~Musette, and R.~Conte,
12 pages,
\textit{Bilinear integrable systems -
 from classical to quantum, continuous to discrete},
ed.~P.~van Moerbeke (Kluwer, Dordrecht, 2004).
nlin.SI/0405034.

\bibitem{Woj1985} S.~Wojciechowski,
Physica Scripta {\bf 31} (1985) 433--438.

\end{thebibliography}
\end{document}